\documentclass[bm,aps,showpacs,amsfonts,amssymb]{revtex4}  

\usepackage{graphicx}
\usepackage{natbib}

\usepackage{color}
\newcommand\beqa{\begin{eqnarray}}
\newcommand\eeqa{\end{eqnarray}}
\newcommand\n{\nonumber\\}
\begin{document}

{~}

\title{
General Kaluza-Klein black holes with all six independent charges \\
in five-dimensional minimal supergravity 

}
\vspace{2cm}
\author{Shinya Tomizawa\footnote{E-mail:tomizawa@mx.ibaraki.ac.jp} and Shun'ya Mizoguchi\footnote{E-mail:mizoguch@post.kek.jp}}
\vspace{2cm}
\affiliation{
${}^* $ University Educational Center, Ibaraki University, Mito, Ibaraki, 310-8512, Japan,\\
${}^\dagger $Theory Center, Institute of Particle and Nuclear Studies,
KEK, Tsukuba, Ibaraki, 305-0801, Japan 
}
\begin{abstract} 
Using the $SL(2,R)$-duality in a dimensionally reduced spacetime in (the bosonic sector of) five-dimensional minimal supergravity, 
we construct general Kaluza-Klein black hole solutions which carry six independent charges, its mass, angular momentum along four dimensions, electric and magnetic charges of the Maxwell fields in addition to Kaluza-Klein electric and magnetic monopole charges.
\end{abstract}

\preprint{KEK-TH 1466}
\date{\today}
\maketitle
\section{Introduction}

In string theory and various related contexts, higher dimensional black holes have played an important role. 
In particular, physics of black holes in the five-dimensional Einstein-Maxwell-Chern-Simons (EMCS) theory has recently been the subject of increased attention, as the EMCS theory describes the bosonic sector of five-dimensional minimal supergravity, a sub-sector of 
a low-energy limit of string theory.
The dimensional reduction of minimal supergravity to four dimensions yields two Maxwell fields, a massless axion and a dilaton, all coupled to gravity~\cite{CN}, where 
(as was shown in Ref.~\cite{MO},) the equations of motions derived from the dimensional reduction are invariant under the action of a global $SL(2,R)$ group, by which the Maxwell fields are related to the Kaluza-Klein electromagnetic fields.
This so-called $SL(2,R)$-duality enables us to generate a new solution in (the bosonic sector of) five-dimensional minimal supergravity by stating from a certain known solution in the same theory.

\medskip
\begin{table}
\begin{center}
\begin{tabular}{l|llllll}\hline
{\sl Solutions in $D=5$ minimal supergravity} & {\ }$M$ & $J$ & $Q$ & $P$ & $q$ & $p$ \\ \hline
{\sl Gaiotto {\it et al.}}~\cite{Gaiotto} &\  yes${}^\dagger $ & no & yes & yes${}^\dagger $ & yes${}^\dagger $ & no \\ 
{\sl Elvang {\it et al.}}~\cite{Elvang3} &\ yes${}^\dagger $ & no & {yes} & {yes}${}^\dagger $ & {yes}${}^\dagger $ & {yes}${}^\dagger $ \\
{\sl Ishihara-Matsuno}~\cite{IM} &\  yes & no & no & yes & yes & no \\
{\sl Nakagawa {\it et al.}}~\cite{NIMT} {\ }&\ yes &  no & yes${}^\dagger $ & yes${}^\dagger $ & yes${}^\dagger $ & yes${}^\dagger $ \\
{\sl Tomizawa {\it et al.}}~\cite{TIMN}  &\  yes & no & {yes} & yes & yes & yes \\
{\sl Tomizawa {\it et al.}}~\cite{TYM} &\  yes & {yes} & no & yes & yes & no \\
{\sl Compere {\it et al.}}~\cite{CBSV} &\  yes & {yes} & yes & no & yes & yes \\
{\sl Mizoguchi-Tomizawa}~\cite{MT} &\  {yes} & {yes} & {yes}${}^\dagger $ & {yes}${}^\dagger $ & {yes}${}^\dagger $ & {yes}${}^\dagger $ \\ \hline
\end{tabular}
\caption[smallcaption]{Classification of Kaluza-Klein black holes in five-dimensional minimal supergravity: 
The six charges, $M$, $J$, $Q$, $P$, $q$ and $p$ denote, respectively, their 
mass, angular momentum, Kaluza-Klein electric charge, Kaluza-Klein magnetic charge, 
electric charge and magnetic charge.  Here the charges with a dagger `` ${}\dagger{}$ " for each solution are not independent but related by a certain constraint. }
\label{list:KK}
\end{center}
\end{table}

\medskip
Known Kaluza-Klein black hole solutions 
of five-dimensional minimal supergravity are summarized in TABLE~\ref{list:KK}, where 
they are classified by their conserved charges 
(from the four-dimensional perspective) :
mass, angular momentum, Kaluza-Klein electric/magnetic charges and electric/magnetic charges of the Maxwell field. 
As shown in the list, the most general black hole solutions 
with full six independent charges, 
which are expected to exist \cite{Tomizawa:2010xj}, 
have not been discovered so far. 
The aim of this paper is
to present such exact solutions describing general Kaluza-Klein black holes with full six charges 
obtained by using our framework~\cite{MT,MT2} of the $SL(2,R)$-duality.

\medskip
In our previous work~\cite{MT}, applying the $SL(2,R)$-duality symmetry to the Rasheed solutions~\cite{Rasheed}, which are known to describe dyonic rotating black holes (from the four-dimensional point of view) of five-dimensional pure gravity, we obtained six-charge rotating Kaluza-Klein black hole solutions in five-dimensional minimal supergravity. 
However, it turns out that four (Kaluza-Klein  electric/magnetic charges and electric/magnetic charges of the Maxwell field
) of these conserved charges are related by a constraint, namely, these parameters 
are not wholly independent. 
In this paper, we alternatively 
use as a seed solution the boosted rotating (electrically/magnetically charged) black string solutions with five parameters obtained in Ref.~\cite{CBSV}. 
As is well known, the boost along the fifth dimension yields an (Kaluza-Klein's) electric charge in the dimensionally reduced four-dimensional theory. 
Therefore, our starting-point solutions have five charges (mass, angular momentum along four dimensions, the Kaluza-Klein electric charge and  electric/magnetic charges of 
the Maxwell field).
The $SL(2,R)$-duality transformation then adds the missing Kaluza-Klein monopole charge 
to the seed solution.

\medskip
The remainder of this paper is organized as follows: 
In the next section, we present the metric and gauge potential $1$-form of the 
Maxwell field of the seed solution, which is our starting point. 
In Section III, by acting the $SL(2,R)$ transformation on the seed solution, we derive the most general Kaluza-Klein black hole solutions in the above sense, and write down the metric and Maxwell fields.
In Section IV, we show that our solution describes rotating black holes with six conserved charges (mass, angular momentum, Kaluza-Klein electric/magnetic charges and electric/magnetic charges of the Maxwell field) in the dimensionally reduced four-dimensional theory.
In Section V, we discuss limits of our solution to some known ones.
Section VI is devoted to summarizing our results.
The new solution contains complicated polynomials of $r$ (the radial coordinate) 
and $x=\cos \theta$ (with $\theta$ being an angle coordinate), and their coefficients 
are collected in Appendix A. Finally, Appendix B is a brief summary of the $SL(2,R)$ duality.

\section{Seed solution}
In this paper, we choose the black string solutions with five independent parameters as a seed solution in Ref~\cite{CBSV}, whose metric and gauge potential $1$-form are given by, respectively,
\begin{eqnarray}
ds^2&=&\frac{\Sigma}{\xi(F_1+\Delta_2)^2}\left(dx^5+B_\mu dx^\mu\right)^2\nonumber\\
&&+\frac{\xi^{1/2}(F_1+\Delta_2)}{\Sigma^{1/2}}\left[-\frac{\xi^{1/2}\Delta_2}{\Sigma^{1/2}}\left(dt+\omega d\phi\right)^2+\frac{\Sigma^{1/2}}{\xi^{1/2}}\left(\frac{\Delta (1-x^2)}{\Delta_2}d\phi^2+\frac{dr^2}{\Delta}+\frac{dx^2}{1-x^2}\right)\right],\label{eq:seedmetric}
\end{eqnarray}
\begin{eqnarray}
A&=&-\frac{\sqrt{3}(c_gF_3+s_gF_2)}{F_1+\Delta_2}dt+\biggl[\frac{2\sqrt{3}mc_d^2}{\Delta_2}\biggl\{-2s_bc_bf\Delta x+as_d(2ms_b^2(6c_b^2c_d^2f^2-1)+(c_b^2+s_b^2)r)(1-x^2)\biggr\}\nonumber\\
&&-\frac{\sqrt{3}}{F_1+\Delta_2}\left\{\frac{4mas_bc_bc_d^3f(r+2s_b^2m)}{\Delta_2}(1-x^2)+F_3\omega_\phi\right\} \biggr]d\phi-\frac{\sqrt{3}(c_gF_2+s_gF_3)}{F_1+\Delta_2}dx^5,\label{eq:seedA}
\end{eqnarray}
where 
\begin{eqnarray}
B_\mu dx^\mu&=&\frac{-c_gs_g\Delta_2(F_1+\Delta_2)^3+(s_gk+c_g\xi)(c_gk+s_g\xi)}{\Sigma}dt\nonumber\\
  &&+\frac{k\omega_\phi(ks_g+c_g\xi)-s_g\Delta_2(F_1+\Delta_2)^3\omega_\phi+\frac{(s_gk+c_g\xi)(4mac_bs_bc_d^3(r+2ms_b^2))(1-x^2)}{\sqrt{1+3c_d^2}\Delta_2}}{\Sigma}d\phi,
\end{eqnarray}
\begin{eqnarray}
\omega=c_g\omega_\phi+s_g\frac{-4ma c_bs_bc_d^3(r+2m s_b^2)}{\sqrt{1+3c_d^2}\Delta_2}(1-x^2),
\end{eqnarray}
\begin{eqnarray}
\omega_\phi=-2mac_d^3\frac{1-x^2}{\Delta_2}\left[2ms_b^2\left(c_b^2+s_b^2-\frac{4c_b^2}{1+3c_d^2}\right)+r(c_b^2+s_b^2)\right],
\end{eqnarray}
\begin{eqnarray}
\Sigma=(s_gk+c_g\xi)^2-s_g^2\Delta_2(F_1+\Delta_2)^3,
\end{eqnarray}
\begin{eqnarray}
\Delta=r^2-2mr+a^2,\quad \Delta_2=r^2-2mr+a^2x^2,
\end{eqnarray}
\begin{eqnarray}
\xi=(F_4+\Delta_2)(F_1+\Delta_2)-F_2^2,\quad k=F_5(F_1+\Delta_2)-F_2F_3,
\end{eqnarray}
\begin{eqnarray}
&&F_1=2mc_d^2\left[2ms_b^2(s_b^2+s_d^2c_b^2f^2)+(c_b^2+s_b^2)r+2as_ds_bc_bfx\right],\\
&&F_2=2ms_dc_d\left[2ms_bc_bf(-1+(c_b^2+s_b^2)c_d^2)+2s_bc_bf r+as_d(c_b^2+s_b^2)x\right],\\
&&F_3=2mc_d\left[2ms_ds_b^2(c_b^2(1+c_d^2)f^2-s_b^2)-s_d(c_b^2+s_b^2)r+2as_bc_bfx\right],\\
&&F_4=2m\left[2m((c_b^2s_d^2+s_b^2c_d^2)^2+s_d^2s_b^2c_b^2f^2)+(c_b^2+s_b^2)(c_d^2+s_d^2)r-2as_bc_bs_dfx\right],\\
&&F_5=2m\left[2ms_bc_bf(-1+(c_b^2+s_b^2)c_d^4)+2s_bc_bfr-as_d^2(c_b^2+s_b^2)x\right],
\end{eqnarray}
\begin{eqnarray}
&&s_b=\sinh b,\ c_b=\cosh b,\ s_d=\sinh d,\ c_d=\cosh d,\ s_g=\sinh g,\ c_g=\cosh g,\\
&&f=\frac{1}{\sqrt{1+3c_d^2}}.
\end{eqnarray}
Here, we would like to note that our definition for the function $\Sigma$ is different from one in Ref.~\cite{CBSV}.
As will be seen later, the familiar parameters $m$ and $a$ denote the mass and rotational parameter, respectively, and three parameters, $b$ and $d$ and $g$, correspond to the magnetic charge, electric charge of the Maxwell $U(1)$ gauge field and electric charge of the Kaluza-Klein $U(1)$ gauge field, respectively.

\section{Transformation}
Now, in order to construct six-charge solutions, we apply the $SL(2,R)$-duality transformation to the above solution. 
Some necessary transformation formula developed in our previous work~\cite{MT} are briefly  summarized in Appendix B.
From Eqs.~(\ref{eq:seedmetric}) and (\ref{eq:seedA}), one can read off the dilaton and axion for the seed as
\begin{eqnarray}
\rho=\frac{\Sigma^{1/2}}{\sqrt{\xi}(F_1+\Delta_2)},\quad
A_5=-\sqrt{3}\frac{s_gF_3+c_gF_2}{F_1+\Delta_2}.
\end{eqnarray}

Therefore, from Eqs.~(\ref{eq:newrho}) and (\ref{eq:newA5}) in Appendix B, the dilaton and axion fields for the transformed solutions are written as, respectively,  
\begin{eqnarray}
\rho_{new}=\frac{\Sigma^{1/2} \xi^{1/2} (F_1+\Delta_2)}{\Pi^2+\beta^2 \Sigma},
\end{eqnarray}
\begin{eqnarray}
A_5^{new}=\sqrt{3}\xi^{1/2}\frac{\Pi[\alpha(F_1+\Delta_2)-(c F_2+sF_3)]+\beta\Sigma}{\Pi^2+\beta^2 \Sigma},
\end{eqnarray}
where the function $\Pi$ is
\begin{eqnarray}
\Pi=\xi^{1/2}[\gamma(F_1+\Delta_2)-\beta(cF_2+sF_3)].
\end{eqnarray}

\medskip
On the other hand, from Eqs.~(\ref{eq:Bnew1}) and (\ref{eq:Anew1}) in Appendix B,
the gauge potential $1$-forms of Kaluza-Klein's and Maxwell's $U(1)$ gauge fields  for the transformed solutions are written as
\begin{eqnarray}
B_\mu^{new}=\sqrt{3}\beta^2\gamma \tilde A_\mu+\left(\gamma^3+\sqrt{3} \beta\gamma^2 A_5\right)B_\mu-\sqrt{3} \beta\gamma^2 A_\mu+\beta^3 \tilde B_\mu,
\end{eqnarray}
\begin{eqnarray}
A_\mu^{new}&=&\left[\sqrt{3}\beta^2\gamma A_5^{new}-\beta(2+3\alpha\beta) \right]\tilde A_\mu+\left[-\sqrt{3}\alpha\gamma^2+\gamma^3 A_5^{new}-A_5((1+4\alpha\beta+3\alpha^2\beta^2)-\sqrt{3}\beta\gamma^2 A_5^{new}))\right]B_\mu\nonumber\\
&&+\left[(1+4\alpha\beta+3\alpha^2\beta^2)-\sqrt{3}\beta\gamma^2 A_5^{new})\right] A_\mu+\left[-\sqrt{3}\beta^2+\beta^3 A_5^{new}\right]\tilde B_\mu,
\end{eqnarray}
where the $1$-forms $\tilde B_\mu dx^\mu\ (\mu=t,\phi)$ and $\tilde A_\mu dx^\mu\ (\mu=t,\phi)$ can be 
obtained by integrating  Eqs. (\ref{eq:tB}) and (\ref{eq:tildeA1}) in Appendix B, and they  are explicitly written as
\begin{eqnarray}
\tilde B_t&=&\frac{(a_1+a_2x)r^2+(a_3+a_4x)r+a_5+a_6x+a_7x^2+a_8x^3}{\Gamma},\\ \label{eq:tildeBt}
\tilde B_\phi&=&c_1+c_2x+(1-x^2)\frac{b_1r^3+(b_2+b_3x)r^2+(b_4+b_5x+b_6x^2)r+b_7+b_8x+b_9x^2+b_{10}x^3}{\Gamma},\\ \label{eq:tildeBphi}
\tilde A_t&=&\frac{p_1r^3+(p_2+p_3x)r^2+(p_4+p_5x+p_6x^2)r+p_7+p_8x+p_9x^2+p_{10}x^3}{\Gamma},\\ \label{eq:tildeAt}
\tilde A_\phi&=&r_1+r_2x+(1-x^2)\frac{q_1r^3+(q_2+q_3x)r^2+(q_4+q_5x+q_6x^2)r+q_7+q_8x+q_9x^2+q_{10}x^3}{\Gamma}, \label{eq:tildeAphi}
\end{eqnarray}
where the function $\Gamma$ is
\begin{eqnarray}
\Gamma=\frac{-s^2\Delta_2(F_1+\Delta_2)^3+(c_g\xi+s_gk)^2}{\xi},
\end{eqnarray}
and the constants $a_1,\cdots,a_8$, $b_1,\cdots,b_{10}$, $c_1,c_2$, $p_1,\cdots,p_{10}$, $q_1,\cdots,q_{10}$, $r_1,r_2$ are related to the five-parameters $m,a,b,d$ and $g$ only (the explicit forms are given in Appendix A). Note that the constant $c_1$ can be set to be zero by a coordinate transformation.

\section{Most general solutions}
From the previous section, we can get six-charge Kaluza-Klein solutions in the same theory, whose metric and gauge potential $1$-form are written as, respectively,
\begin{eqnarray}
ds^2&&=\frac{\Sigma \xi (F_1+\Delta_2)^2}{(\Pi^2+\beta^2 \Sigma)^2}\left[dx^5+\left\{ \sqrt{3}\beta^2\gamma \tilde A_\mu+\left(\gamma^3+\sqrt{3} \beta\gamma^2 A_5\right)B_\mu-\sqrt{3} \beta\gamma^2 A_\mu+\beta^3 \tilde B_\mu  \right\} dx^\mu\right]^2\nonumber\\
&&+\frac{\Pi^2+\beta^2 \Sigma}{\Sigma^{1/2} \xi^{1/2} (F_1+\Delta_2)}\left[-\frac{\xi^{1/2}\Delta_2}{\Sigma^{1/2}}\left(dt+\omega d\phi\right)^2+\frac{\Sigma^{1/2}}{\xi^{1/2}}\left(\frac{\Delta}{\Delta_2}(1-x^2)d\phi^2+\frac{dr^2}{\Delta}+\frac{dx^2}{1-x^2}\right)\right],
\end{eqnarray}
\begin{eqnarray}
A^{new}&=&\biggl[ \left\{\sqrt{3}\beta^2\gamma A_5^{new}-\beta(2+3\alpha\beta) \right\}\tilde A_\mu+\left\{-\sqrt{3}\alpha\gamma^2+\gamma^3 A_5^{new}-A_5((1+4\alpha\beta+3\alpha^2\beta^2)-\sqrt{3}\beta\gamma^2 A_5^{new}))\right\} B_\mu\nonumber\\
&&+\left\{(1+4\alpha\beta+3\alpha^2\beta^2)-\sqrt{3}\beta\gamma^2 A_5^{new}\right\} A_\mu+\left\{-\sqrt{3}\beta^2+\beta^3 A_5^{new}\right\}\tilde B_\mu \biggr] dx^\mu \nonumber\\
&&+\left[\sqrt{3}\xi^{1/2}\frac{\Pi[\alpha(F_1+\Delta_2)-(c_g F_2+s_gF_3)]+\beta\Sigma}{\Pi^2+\beta^2 \Sigma} \right]dx^5.
\end{eqnarray}

\section{charges}
In this theory, the electric/magnetic charges ($Q/P$) of the Kaluza-Klein $U(1)$ field and electric/magnetic charges ($q/p$) of the Maxwell field are defined by, respectively,
\begin{eqnarray}
&&Q=\frac{1}{8\pi}\int_{S^2}{\cal H}^B, \ P=\frac{1}{8\pi}\int_{S^2} {\cal B},
\end{eqnarray}
\begin{eqnarray}
&&q=\frac{1}{8\pi}\int_{S^2}\tilde {\cal A}, \ p=\frac{1}{8\pi}\int_{S^2}{\cal F}'.
\end{eqnarray}
where $S^2$ denotes any closed two-surfaces surrounding the black hole. The two form fields ${\cal H}^B$ and ${\cal B}$ are defined by ${\cal H}^B:=\frac{1}{2}H^B_{\mu\nu}dx^\mu\wedge dx^\nu$ and ${\cal B}:=\frac{1}{2}B_{\mu\nu}dx^\mu\wedge dx^\nu$, respectively. Similarly, where $\tilde{\cal A}:=\frac{1}{2}\tilde A_{\mu\nu}dx^\mu\wedge dx^\nu$ and ${\cal F}':=\frac{1}{2}F'_{\mu\nu}dx^\mu\wedge dx^\nu$.
The four charges for the new solution are related to those of seed by
\begin{eqnarray}
\left(
\begin{array}{c}
q^{new}\\
P^{new}\\
-p^{new}\\
Q^{new}\\
\end{array}
\right)=
\left(
\begin{array}{cccc}
 1+3 \alpha  \beta  & \sqrt{3} \alpha ^2 (1+\alpha  \beta ) & \alpha  (2+3 \alpha  \beta ) & \sqrt{3} \beta  \\
 \sqrt{3} \beta ^2 (1+\alpha  \beta ) & (1+\alpha  \beta )^3 & \sqrt{3} \beta  (1+\alpha  \beta )^2 & \beta ^3 \\
 \beta  (2+3 \alpha  \beta ) & \sqrt{3} \alpha  (1+\alpha  \beta )^2 & 1+4 \alpha  \beta +3 \alpha ^2 \beta ^2 & \sqrt{3} \beta ^2 \\
 \sqrt{3} \alpha  & \alpha ^3 & \sqrt{3} \alpha ^2 & 1
\end{array}
\right)\left(
\begin{array}{c}
q\\
0\\
-p\\
Q\\
\end{array}
\right),
\end{eqnarray}
where $Q$ and $q/p$ are the electric charge for Kaluza-Klein $U(1)$ field and electric/magnetic charge for Maxwell field for the seed solution can be explicitly written as
\begin{eqnarray}
Q&=&m\left[(1+3s_d^2)(c_b^2+s_b^2)s_gc_g+2s_bc_b(c_g^2+s_g^2)f\right],\\
q&=&\sqrt{3}ms_dc_d\left[(c_b^2+s_b^2)c_g-2s_bc_bs_gf\right],\\
p&=&2\sqrt{3}ms_bc_bc_d^2f.
\end{eqnarray}
Note that for the seed (boosted black string), the Kaluza-Klein's magnetic monopole charge $P$ vanishes. 
One of the two parameters $\alpha$ and $\beta$ corresponds to adding a constant to $A_5$, which means that gauge transformation for the potential $1$-form $A_M$. 
Therefore, the remaining one corresponds to adding a physical degree of freedom, i.e., a Kaluza-Klein monopole charge.


\medskip
In our previous work~\cite{MT}, 
we derived a different six-charge solution 
starting from a different seed solution, but  the charges of that solution were 
not wholly independent, but 
four of them $(P,Q,p,q)$ were related by a certain constraint. 
Here we would like to confirm whether or not these conserved charges are actually independent.
As is easily verified, the following Jacobian does not vanish
\begin{eqnarray}
\frac{\partial(Q^{new},P^{new},q^{new},p^{new})}{\partial(b,d,g,\beta)},
\end{eqnarray}
which therefore means that these charges are independent of one another. 
Also, it is clear that the mass $M$ and angular momentum $J$ are not related to these four charges.

\section{Some limits}

Here, we study some limits to known Kaluza-Klein black hole solutions by taking some parameter limits in our solutions.
First, we take the limit of $\alpha=-1, \beta=1,\  a=0,\ d=0$. Defining the parameters $(\varrho_\pm,\varrho_0)$ by
\begin{eqnarray}
\varrho_-=2ms_b^2,\ \varrho_+=2m c_b^2,\ \varrho_0=2ms_g(s_g+2s_b^2s_g+2s_bc_bc_g),
\end{eqnarray}
and the coordinates $(\varrho,\theta,\psi)$ by
\begin{eqnarray}
r=\varrho+m(1-c_b^2-s_b^2),\ x=-\cos\theta, \ x^5=2m(s_gc_g+2s_b^2s_gc_g+s_bc_b(c_g+s_g))\psi,
\end{eqnarray}
one can obtain the metric: 
\begin{eqnarray}
ds^2&=&-\frac{(\varrho-\varrho_+)(\varrho-\varrho_-)}{\varrho^2}dt^2+\frac{\varrho(\varrho+\varrho_0)}{(\varrho-\varrho_+)(\varrho-\varrho_-)}d\varrho^2+\varrho(\varrho+\varrho_0)(d\theta^2+\sin^2\theta d\phi^2)\nonumber\\
    & &+\frac{\rho(\varrho_++\varrho_0)(\varrho_-+\varrho_0)}{\varrho+\varrho_0}(d\psi+\cos\theta d\phi)^2.
\end{eqnarray}
This is the metric of the Ishihara-Matsuno solutions~\cite{IM}.

\medskip

Next, we identify the coordinate and the parameters as
$\varrho:=r-2ms_b^2$, $\gamma:=-b$, $\beta:=b-g$, 
$M_k:=m$. 
Then, the metric can be written as
\begin{eqnarray}
ds^2=\rho^2(dt+B_tdt+ B_\phi d\phi)^2+\rho^{-1}ds^2_{(4)},
\end{eqnarray}
where 
\begin{eqnarray}
\rho^2=\frac{\varrho^3(\varrho+2M_k(s_\beta^2-s_\gamma^2))+2a^2x^2[\varrho^2+M_k(s_\beta^2-s_\gamma^2)-2M_k^2s_\gamma^2c_\gamma^2(s_\gamma^2+2s_\beta c_\beta s_\gamma c_\gamma+s_\beta^2c_\gamma^2)]+a^4x^4}{\varrho^2(\varrho-2s_\gamma^2M_k+2s_\beta^2M_k)^2},
\end{eqnarray}
\begin{eqnarray*}
B_t=-2M_kax\frac{[\varrho+2M_k(s_\beta^2-s_\gamma^2)][-2M_ks_\gamma^2c_\gamma^2(c_\beta s_\gamma+s_\beta s_\gamma)+(c_\beta s_\gamma^3+s_\beta c_\gamma+s_\beta s_\gamma^2c_\gamma)\varrho]+M_ka^2x^2(c_\beta s_\gamma^3+s_\beta c_\gamma+s_\beta s_\gamma^2c_\gamma)}{\varrho^3(\varrho+2M_k(s_\beta^2-s_\gamma^2))+2a^2x^2[\varrho^2+M_k(s_\beta^2-s_\gamma^2)-2M_k^2s_\gamma^2c_\gamma^2(s_\gamma^2+2s_\beta c_\beta s_\gamma c_\gamma+s_\beta^2c_\gamma^2)]+a^4x^4},
\end{eqnarray*}
\begin{eqnarray}
B_\phi=2M_ks_\beta c_\beta x\frac{(\varrho-2s_\gamma^2 M_k)(\varrho-2c_\gamma^2 M_k)+a^2}{(\varrho-2s_\gamma^2 M_k)(\varrho-2c_\gamma^2 M_k)+a^2x^2}+B_t\omega^0{}_\phi,
\end{eqnarray}
\begin{eqnarray}
ds^2_{(4)}&=&-\frac{\varrho^2-2M_k(c_\gamma^2+s_\gamma^2)\varrho+4M_k^2c_\gamma^2s_\gamma^2+a^2x^2}{\rho^2[\varrho^2+2M_k(s_\beta^2-s_\gamma^2)\varrho+a^2x^2]}\left(dt+\omega^0{}_\phi d\phi\right)^2\nonumber\\
&&+\rho\frac{[\varrho^2+2M_k(s_\beta^2-s_\gamma^2)\varrho+a^2x^2][(\varrho-2s_\gamma^2 M_k)(\varrho-2c_\gamma^2 M_k)+a^2]}{(\varrho-2s_\gamma^2 M_k)(\varrho-2c_\gamma^2 M_k)+a^2x^2}d\phi^2\\
&&+\rho\varrho[\varrho+2M_k(s_\beta^2-s_\gamma^2)] \left[\frac{d\varrho^2}{(\varrho-2s_\gamma^2 M_k)(\varrho-2c_\gamma^2 M_k)+a^2}+\frac{dx^2}{1-x^2}\right],\nonumber
\end{eqnarray}
\begin{eqnarray}
\omega^0{}_\phi=-2M_ka\frac{-\varrho(s_\beta s_\gamma^3+c_\beta c_\gamma^3)+2M_ks_\gamma^2c_\gamma^2(s_\beta s_\gamma+c_\beta s_\gamma)}{(\varrho-2s_\gamma^2 M_k)(\varrho-2c_\gamma^2 M_k)+a^2x^2}.
\end{eqnarray}
This exactly coincides with the metric of the {\sl rotating} Ishihara-Matsuno solutions in Ref.~\cite{TYM}.


\section{Summary}

In this paper, using the $SL(2,R)$-duality transformation that the reduced Lagrangian possesses upon reduction to four dimensions, we have succeeded in constructing general Kaluza-Klein black hole solutions in (the bosonic sector of) five-dimensional minimal supergravity, where we have used the electrically/magnetically charged boosted black string solution as a seed solution.
Our solutions are the most general ones in the sense that in that theory, from a four-dimensional point of view, such a class of regular black hole solutions can be specified by six independent charges, its mass, angular momentum along four dimensions, electric and magnetic charges of the Maxwell fields in addition to the Kaluza-Klein electric and magnetic monopole charges.
  From the five-dimensional point of view, like known Kaluza-Klein charged black hole solutions, the black hole spacetime has two horizons, the outer and inner horizons, and although the cross-section geometry of the outer horizon is of $S^3$, at large distances the spacetime behaves effectively as a four-dimensional spacetime, which is due to the existence of a Kaluza-Klein monopole charge. 
\medskip

The present metric form of our solutions is considerably lengthy and complicated, 
which, as a result, makes us difficult to analyze the physical properties of our solutions.  
To do so, it should be written in terms of some physical parameters such as $(M,J,Q,P,q,p)$ rather than the transformation parameters $\alpha,\beta,b,d,g$.
As a future work, we would like to present it in a more compact and physically clearer form.

\section*{Acknowledgments} 
We would like to thank H. Kodama for valuable discussions and comments.
The work of S.~M. and S.~T. is supported by 
Grant-in-Aid
for Scientific Research  
(A) \#22244430-0007, and S.~M. 
is also by (C) \#20540287-H20  
from
The Ministry of Education, Culture, Sports, Science
and Technology of Japan.

\appendix
\section{Coefficients}
The coefficients $a_1,\cdots,a_8,b_1,\cdots,b_{10},c_1,c_2,p_1,\cdots,p_{10},q_1,\cdots,q_{10},r_1,r_2$ in Eqs.(\ref{eq:tildeBt})-(\ref{eq:tildeAphi}) are given by, respectively, 
\begin{eqnarray}
a_1&=&-12 m^2s_b s_d c_d^3 f\left[c_b \left( s_b^2+c_b^2\right) s_g-2 s_b c_b^2 c_gf\right],\\
a_2&=&a_8a^{-2}=-2m a  c_d^3 \left[\left( s_b^2+c_b^2\right) s_g+2 s_b c_b c_gf\right],\\
a_3&=&-8 m^3 s_d c_d^3\biggl[s_d^2 s_g^2 c_g+s_b c_bs_g \left(s_d^2+3 c_d^2 s_g^2\right)f+4 s_b^6 c_g \left(-3-4 s_d^2+\left(3+5 s_d^2+6 s_d^4\right) s_g^2\right)f^2\nonumber\\
&&+6 s_b^4c_g \left(-2-4 s_d^2+\left(3+5 s_d^2+6 s_d^4\right) s_g^2\right)f^2+2 s_b^3 c_b s_g \left(-6+9 s_d^2+6 s_d^4+2 \left(6+19 s_d^2+9 s_d^4\right) s_g^2\right)f^3\nonumber\\
&&+2 s_b^2 c_g \left(-4 s_d^2+3 \left(1+3 \left(s_d^2+s_d^4\right)\right) s_g^2\right)f^2+4 s_b^5 c_b s_g \left(3+6 s_g^2+s_d^4 \left(3+9 s_g^2\right)+s_d^2 \left(9+19 s_g^2\right)\right)f^3\biggr],\\
a_4&=&-4 m^2a c_d^3\biggl[6 s_b c_b \left(4+7 s_d^2+3 s_d^4\right) s_g^2 c_gf^3+4 s_b^3 c_b c_g \left(1+3 c_d^2 s_g^2\right)f+ s_g \left(s_g^2+3 s_d^2 c_g^2\right)\nonumber\\
&&+4 s_b^4s_g \left(5+6 s_g^2+3 s_d^2 \left(5+3 s_d^2\right) c_g^2\right)f^2+6 s_b^2 s_g \left(2+4 s_g^2+6 s_d^4 c_g^2+s_d^2 \left(9+10 s_g^2\right)\right)f^2\biggr],\\
a_5&=&16 m^4 s_b^2 s_dc_d^3 \biggl[12 s_b^5 c_b c_d^4s_gf^3+12 s_b^6 c_d^4 \left(2+3 s_d^2\right) c_gf^4+s_b c_b s_d^2\left(17+9 s_d^2\right) s_g c_g^2f^3+3 s_d^4c_g^3f^2\nonumber\\
&&+2 s_b^3 c_b s_g \left(6 \left(2+s_g^2\right)+3 s_d^4 \left(4+3 s_g^2\right)+s_d^2 \left(27+19 s_g^2\right)\right)f^3+2 s_b^4 c_g (12 c_g^2+18 s_d^6 \left(2+s_g^2\right)\nonumber\\
&&+s_d^2 \left(59+29 s_g^2\right)+s_d^4 \left(84+39 s_g^2\right))f^4+s_b^2 c_g \left(24 s_g^2+9 s_d^6 \left(5+4 s_g^2\right)+s_d^2 \left(34+42 s_g^2\right)+s_d^4 \left(84+66 s_g^2\right)\right)f^4\biggr],\\
a_6&=&-8 m^3a s_b^2 c_d^3\biggl[3 s_d^2 \left(4+15 s_d^2+9 s_d^4\right) s_g c_g^2f^4+2 s_b c_b c_g \left(3 s_d^2+\left(4+21 s_d^2+9 s_d^4\right) s_g^2\right)f^3
+4 s_b^3 c_bc_g (2+3 s_d^2\nonumber\\
&&+\left(8+21 s_d^2+9 s_d^4\right) s_g^2)f^3+4 s_b^4 s_g \left(3+4 s_g^2+9 s_d^2 c_d^2 c_g^2\right)f^2+4 s_b^2 s_g \left(2+3 s_g^2+3 s_d^2 \left(2+3 s_d^2\right) c_g^2\right)f^2\biggr],\\
a_7&=&4 m^2a^2 s_d c_d^3 \biggl[s_d^2 c_g \left(1+2 s_g^2\right)+s_b c_b  s_g \left(3+4 s_d^2+6 c_d^2 s_g^2\right)f+2 s_b^3 c_b  s_g \left(3+4 s_d^2+6 c_d^2 s_g^2\right)f\nonumber\\
&&+2 s_b^2 c_g \left(3+8 s_d^2+6 s_d^4+6 \left(1+3 s_d^2+2 s_d^4\right) s_g^2\right)f^2+2 s_b^4 c_g \left(3+8 s_d^2+6 s_d^4+6 \left(1+3 s_d^2+2 s_d^4\right) s_g^2\right)f^2\biggr],
\end{eqnarray}
\begin{eqnarray}
b_1&=&b_6a^{-2}=2 m as_d \left[-2 s_b c_b \left(3+s_d^2\right) s_g c_gf-\left(c_b^2+ s_b^2\right) s_d^2 \left(1+2 s_g^2\right)\right],\\
b_2&=&-4  m^2a s_d \biggl[4 s_b^3 c_b s_g c_g \left(6+13 s_d^2+3 s_d^4+\left(3+14 s_d^2+3 s_d^4\right) s_g^2\right)f\nonumber\\
 &&+2 s_b^4 \left(-3+10 s_d^2+33 s_d^4+18 s_d^6+2 \left(6+22 s_d^2+51 s_d^4+27 s_d^6\right) s_g^2+4 \left(3+5 s_d^2+15 s_d^4+9 s_d^6\right) s_g^4\right)f^2\nonumber\\
&&+s_b^2 \left(-6+8 s_d^2+57 s_d^4+36 s_d^6+2 \left(12+32 s_d^2+93 s_d^4+54 s_d^6\right) s_g^2+8 \left(3+5 s_d^2+15 s_d^4+9 s_d^6\right) s_g^4\right)f^2\nonumber\\
&&+s_d^2 c_g^2 \left(2 s_g^2+s_d^2 \left(3+6 s_g^2\right)\right)+s_b c_b s_g c_g \left(3+6 s_g^2+6 s_d^4 c_g^2+s_d^2 \left(23+28 s_g^2\right)\right)f\biggr],\\
b_3&=&-2 ma^2 \left[\left(1+3 s_d^2\right) s_g c_g+2 s_b^2 \left(1+3 s_d^2\right) s_g c_g+2 s_b c_b \left(1+2 s_g^2\right)f\right],\\
b_4&=&-8 m^3a s_d f^4\biggl[s_d^2 f^{-4} c_g^2 \left(3 s_d^4+\left(1+3 s_d^2+6 s_d^4\right) s_g^2\right)\nonumber\\
&&+12 s_b^5 c_b c_d^4 f^{-1} s_g c_g \left(7+24 s_d^2+9 s_d^4+2 \left(6+19 s_d^2+9 s_d^4\right) s_g^2\right)\nonumber\\
&&+s_b c_b f^{-3} s_g c_g \left(s_d^2+24 s_d^4+9 s_d^6+\left(3+14 s_d^2+45 s_d^4+18 s_d^6\right) s_g^2\right)\nonumber\\
&&+12 s_b^6 c_d^4 f^{-2} \left(-1+5 s_g^2+6 s_g^4+6 s_d^4 \left(1+3 s_g^2+2 s_g^4\right)+s_d^2 \left(2+11 s_g^2+10 s_g^4\right)\right)\nonumber\\
&&+2 s_b^4 f^{-2} (-6+33 s_g^2+42 s_g^4+54 s_d^8 \left(1+3 s_g^2+2 s_g^4\right)+9 s_d^6 \left(12+41 s_g^2+30 s_g^4\right)\nonumber\\
&&+3 s_d^4 \left(16+101 s_g^2+94 s_g^4\right)+s_d^2 \left(-10+145 s_g^2+178 s_g^4\right))\nonumber\\
&&+4 s_b^3 c_b f^{-1} s_g c_g \left(-3+24 s_g^2+s_d^2 \left(44+121 s_g^2+3 s_d^2 \left(47+39 s_d^2+9 s_d^4+\left(88+71 s_d^2+18 s_d^4\right) s_g^2\right)\right)\right)\nonumber\\
&&+s_b^2 f^{-2} (6 \left(s_g^2+2 s_g^4\right)+54 s_d^8 \left(1+3 s_g^2+2 s_g^4\right)+6 s_d^6 \left(15+53 s_g^2+39 s_g^4\right)+2 s_d^2 \left(-4+43 s_g^2+58 s_g^4\right)\nonumber\\
&&+3 s_d^4 \left(7+66 \left(s_g^2+s_g^4\right)\right))\biggr],\\
b_5&=&4 a^2 m^2f^3 \biggl[f^{-3} s_g c_g \left(3 s_d^2 \left(-1-2 s_d^2+s_d^4\right)+\left(-1-6 s_d^2-9 s_d^4+4 s_d^6\right) s_g^2\right)\nonumber\\
&&+2 s_b^2 f^{-1} s_g c_g \left(3 \left(-2-11 s_d^2-25 s_d^4-4 s_d^6+6 s_d^8\right)+4 \left(-4-9 s_d^2-24 s_d^4-5 s_d^6+6 s_d^8\right) s_g^2\right)\nonumber\\
&&+2 s_b c_b f^{-2} \left(3 s_d^4+3 \left(-1-3 s_d^2+5 s_d^4+s_d^6\right) s_g^2+4 \left(-1-3 s_d^2+3 s_d^4+s_d^6\right) s_g^4\right)\nonumber\\
&&+4 s_b^3 c_b f^{-2} \left(-1+3 s_d^4+\left(-5+3 s_d^2 \left(-3+5 s_d^2+s_d^4\right)\right) s_g^2+4 \left(-1-3 s_d^2+3 s_d^4+s_d^6\right) s_g^4\right)\nonumber\\
&&+4 s_b^4 f^{-1} s_g c_g \left(-5-8 s_g^2-6 s_d^2 \left(4+3 s_g^2\right)+3 s_d^8 \left(3+4 s_g^2\right)-2 s_d^6 \left(3+5 s_g^2\right)-6 s_d^4 \left(7+8 s_g^2\right)\right)\biggr],\\
b_7&=&-16 a m^4 s_d f^5\biggl[s_b c_b s_d^6 f^{-4} \left(9+5 s_d^2\right) s_g c_g^3+s_d^8 f^{-5} c_g^4+4 s_b^7 c_b c_d^6 f^{-2} s_g c_g(3+31 s_d^2+30 s_d^4\nonumber\\
&&+6 \left(3+8 s_d^2+5 s_d^4\right) s_g^2)+s_b^3 c_b s_d^2 f^{-2} s_g c_g(-17+78 s_d^2+318 s_d^4+309 s_d^6+90 s_d^8+(27+150 s_d^2\nonumber\\
&&+357 s_d^4+316 s_d^6+90 s_d^8) s_g^2)+12 s_b^5 c_b c_d^4 f^{-2} s_g c_g(-2+3 s_g^2+15 s_d^6 c_g^2+s_d^2 \left(5+14 s_g^2\right)+s_d^4 \left(26+30 s_g^2\right))\nonumber\\
&&+s_b^2 s_d^4 f^{-3} c_g^2 \left(-3+27 s_g^2+24 s_d^6 c_g^2+3 s_d^2 \left(4+15 s_g^2\right)+s_d^4 \left(41+50 s_g^2\right)\right)\nonumber\\
&&+4 s_b^8 c_d^6 f^{-1}\left(-6+24 s_g^2+36 s_g^4+36 s_d^6 c_g^4+6 s_d^4 c_g^2 \left(7+16 s_g^2\right)+s_d^2 \left(1+92 s_g^2+96 s_g^4\right)\right)\nonumber\\
&&+2 s_b^6 c_d^4 f^{-1}(-12+36 s_g^2+72 s_g^4+144 s_d^8 c_g^4+6 s_d^6 c_g^2 \left(43+70 s_g^2\right)+s_d^2 \left(-35+287 s_g^2+354 s_g^4\right)\nonumber\\
&&+s_d^4 \left(82+605 s_g^2+534 s_g^4\right))+s_b^4 f^{-1}(-24 s_g^2+s_d^2 (-34+142 s_g^2+228 s_g^4+216 s_d^{10} c_g^4+18 s_d^8 c_g^2 \left(41+50 s_g^2\right)\nonumber\\
&&+6 s_d^2 \left(-8+149 s_g^2+164 s_g^4\right)+3 s_d^4 \left(102+661 s_g^2+564 s_g^4\right)+s_d^6 \left(843+2469 s_g^2+1628 s_g^4\right)))\biggr],\\
b_8&=&8 a^2 m^3f^{4} \biggl[s_d^6 \left(-1+3 s_d^2\right) f^{-4} s_g c_g^3+2 s_b c_b s_d^4 f^{-3} c_g^2 \left(2 s_d^2+\left(-3+11 s_d^2+6 s_d^4\right) s_g^2\right)\nonumber\\
&&+s_b^2 s_d^2 f^{-2} s_g c_g\left(-3-6 s_d^2-14 s_d^4+63 s_d^6+54 s_d^8+\left(-15+30 s_d^2+37 s_d^4+78 s_d^6+54 s_d^8\right) s_g^2\right)\nonumber\\
&&+4 s_b^5 c_b c_d^4 f^{-1}\left(-2+s_d^2+12 s_d^4+3 \left(-4-7 s_d^2+21 s_d^4+12 s_d^6\right) s_g^2+6 \left(-2-3 s_d^2+9 s_d^4+6 s_d^6\right) s_g^4\right)\nonumber\\
&&+2 s_b^3 c_b f^{-1}(s_d^2 \left(-3+12 s_d^2+41 s_d^4+24 s_d^6\right)+\left(-4+3s_d^2 \left(-13-5 s_d^2+75 s_d^4+87 s_d^6+24 s_d^8\right)\right) s_g^2\nonumber\\
&&+8 \left(-1+s_d^2 \left(-3+s_d^4 \left(5+3 s_d^2\right)^2\right)\right) s_g^4)+12 s_b^6 c_d^4 f^{-2} s_g c_g\left(-1-2 s_d^2-3 s_d^4-2 s_g^2+6 s_d^6 c_g^2\right)\nonumber\\
&&+4 s_b^4 f^{-2} s_g c_g\left(-2-5 s_g^2+s_d^2 \left(-6-9 s_g^2+15 s_d^2 \left(-1+s_g^2\right)+27 s_d^8 c_g^2+12 s_d^6 \left(3+4 s_g^2\right)+s_d^4 \left(-6+32 s_g^2\right)\right)\right)\biggr],\\
b_9&=&4 a^3 m^2 s_df^2 \biggl[s_d^2 f^{-2} c_g^2 \left(1+\left(2+6 s_d^2\right) s_g^2\right)+2 s_b^4 (3+4 s_d^2+3 s_d^4+2 \left(3+7 s_d^2+30 s_d^4+18 s_d^6\right) s_g^2\nonumber\\
&&+4 \left(3+5 s_d^2+15 s_d^4+9 s_d^6\right) s_g^4)+s_b^2(6+12 s_d^2+9 s_d^4+6 \left(2+6 s_d^2+21 s_d^4+12 s_d^6\right) s_g^2\nonumber\\
&&+8 \left(3+5 s_d^2+15 s_d^4+9 s_d^6\right) s_g^4)+4 s_b^3 c_b f^{-1} s_g c_g \left(3 s_g^2+3 s_d^4 \left(-1+s_g^2\right)+s_d^2 \left(1+14 s_g^2\right)\right)\nonumber\\
&&+s_b c_b f^{-1} s_g c_g \left(3+6 s_g^2+6 s_d^4 \left(-1+s_g^2\right)+s_d^2 \left(3+28 s_g^2\right)\right)\biggr],\\
b_{10}&=&-2 a^4 m \left[\left(1+3 s_d^2\right) s_g c_g+2 s_b^2 \left(1+3 s_d^2\right)s_g c_g+2 s_b c_b \left(1+2 s_g^2\right)f\right],\\
c_2&=&-2 m \left[\left(1+3 s_d^2\right)s_g c_g+2 s_b^2 \left(1+3 s_d^2\right)s_g c_g+2 s_b c_b \left(1+2 s_g^2\right)f\right]
\end{eqnarray}

\begin{eqnarray}
p_1&=&-4 \sqrt{3} m s_b c_b c_d^2f,\\
p_2&=&-4 \sqrt{3} m^2 c_d^2 \biggl[s_d^2 s_g c_g+4 s_b^2 \left(2+5 s_d^2+3 s_d^4\right) s_g c_gf^2\nonumber\\
&&+4 s_b^4 \left(2+5 s_d^2+3 s_d^4\right) s_g c_gf^2+2 s_b^3 c_b c_d^2 \left(3+2 s_g^2\right) f+s_b c_b\left(3 s_d^2+2 c_d^2 s_g^2\right)f\biggr],\\
p_3&=&-2 \sqrt{3}  ma \left(c_b^2+s_b^2\right) s_d c_d^2, \\
p_4&=&-8 \sqrt{3} m^3 c_d^2 \biggl[s_d^4 s_g c_g+6 s_b^2 s_d^2  \left(2+5 s_d^2+3 s_d^4\right) s_g c_gf^{2}+8 s_b^6\left(2+6 s_d^2+7 s_d^4+3 s_d^6\right) s_g c_gf^2\nonumber \\
&&+4 s_b^4 c_d^2\left(4+9 \left(s_d^2+s_d^4\right)\right) s_g c_gf^2+s_b c_b s_d^2 \left(s_d^2+3 c_d^2 s_g^2\right)f+4 s_b^5 c_b c_d^2 \left(6+10 s_d^2+3 s_d^4+\left(8+19 s_d^2+9 s_d^4\right) s_g^2\right)f^3\nonumber\\
&&+2 s_b^3 c_b c_d^2 \left(8 s_g^2+6 s_d^4 \left(1+3 s_g^2\right)+s_d^2 \left(11+32 s_g^2\right)\right)f^3\biggr],\\
p_5&=&-4 \sqrt{3} m^2a s_d c_d^2\left[\left(c_b^2+ s_b^2\right) \left(s_d^2+2 s_b^2 c_d^2\right)+c_d^2 \left(1+12 s_b^2 c_b^2 c_d^2f^2\right) s_g^2\right],\\
p_6&=&-4 \sqrt{3} ma^2  s_b c_b c_d^2f,\\
p_7&=&-16 \sqrt{3} m^4 s_b^3 c_d^4f^4 \biggl[4 s_b^5 c_d^2 \left(8+15 s_d^2+9 s_d^4\right) s_g c_g+s_b s_d^2 \left(20+24 s_d^2+9 s_d^4\right) s_g c_g\nonumber\\
&&+4 s_b^3 \left(8+24 s_d^2+24 s_d^4+9 s_d^6\right) s_g c_g+3 c_b s_d^4 f^{-1} c_g^2+4 s_b^4 c_b f^{-1} \left(2+5 s_d^2+3 s_d^4\right) \left(1+2 s_g^2\right)\nonumber\\
&&+2 s_b^2 c_b f^{-1} \left(4 s_g^2+s_d^4 \left(6+9 s_g^2\right)+s_d^2 \left(5+11 s_g^2\right)\right)\biggr],\\
p_8&=&-8 \sqrt{3} m^3a s_b^2 s_d c_d^4 f^3\biggl[8 s_b c_b^3 s_g c_g+3 s_d^2 f^{-1} c_g^2+4 s_b^2 f^{-1} \left(1+3 s_d^2 c_g^2\right)
 +4 s_b^4 f^{-1} \left(2+s_g^2+3 s_d^2 c_g^2\right)\biggr],\\
p_9&=&4 \sqrt{3}m^2 a^2 c_d^2f^2 \biggl[s_d^4 f^{-2} s_g c_g+4 s_b^2 \left(-1+4 s_d^4+3 s_d^6\right) s_g c_g+4 s_b^4 \left(-1+4 s_d^4+3 s_d^6\right) s_g c_g\nonumber\\
&&+2 s_b^3 c_b c_d^2 f^{-1} \left(-1+2 s_d^2 s_g^2\right)+s_b c_b s_d^2 f^{-1} \left(-1+2 c_d^2 s_g^2\right)\biggr],\\
p_{10}&=&-2 \sqrt{3}m a^3 \left(c_b^2+s_b^2\right) s_d c_d^2, 
\end{eqnarray}

\begin{eqnarray}
q_1&=&-2\sqrt{3} a m c_d f\left[\left(1+2 s_b^2\right) s_d^2 f^{-1} s_g+2 s_b c_b c_g\right],\\
q_2&=&-4 \sqrt{3} a m^2 c_d f^2\biggl[s_d^2 \left(4+15 s_d^2+9 s_d^4\right) s_g c_g^2+2 s_b^4 s_g \left(4+17 s_d^2+37 s_d^4+18 s_d^6+2 \left(2+4 s_d^2+15 s_d^4+9 s_d^6\right) s_g^2\right)\nonumber\\
&&+s_b^2 s_g \left(8+22 s_d^2+65 s_d^4+36 s_d^6+4 \left(2+4 s_d^2+15 s_d^4+9 s_d^6\right) s_g^2\right)+s_b c_b f^{-1} c_g \left(-2 s_d^4+2 s_g^2+s_d^2 \left(3+10 s_g^2\right)\right)\nonumber\\
&&+2 s_b^3 c_b f^{-1} c_g \left(3-2 s_d^4+2 s_g^2+s_d^2 \left(3+10 s_g^2\right)\right)\biggr],\\
q_3&=&-2 \sqrt{3} a^2 m s_d c_df \left[-2 s_b c_b s_g+f^{-1} c_g+2 s_b^2 f^{-1} c_g\right],\\
q_4&=&-8 \sqrt{3} a m^3 c_d f^3\biggl[s_d^4 f^{-1} \left(4+11 s_d^2+6 s_d^4\right) s_g c_g^2+4 s_b^5 c_b c_d^2 c_g \left(6+10 s_d^2-6 s_d^4-9 s_d^6+\left(8+39 s_d^2+33 s_d^4+6 s_d^6\right) s_g^2\right)\nonumber\\
&&+s_b^2 s_d^2 f^{-1} s_g \left(12+31 s_d^2+68 s_d^4+36 s_d^6+2 \left(9+19 s_d^2+35 s_d^4+18 s_d^6\right) s_g^2\right)\nonumber\\
&&+2 s_b^4 f^{-1} s_g \left(8+22 s_d^2+50 s_d^4+78 s_d^6+36 s_d^8+\left(8+31 s_d^2+60 s_d^4+81 s_d^6+36 s_d^8\right) s_g^2\right)\nonumber\\
&&+s_b c_b s_d^2 f^{-2} c_g \left(3 s_g^2+s_d^4 \left(-3+2 s_g^2\right)+s_d^2 \left(1+11 s_g^2\right)\right)+4 s_b^6 c_d^2 f^{-1} s_g \left(4 c_g^2+9 s_d^2 c_g^2+12 s_d^6 c_g^2+s_d^4 \left(20+21 s_g^2\right)\right)\nonumber\\
&&+2 s_b^3 c_b c_g \left(8 s_g^2+s_d^2 \left(11+7 s_d^2-24 s_d^4-18 s_d^6+6 c_d^2 \left(8+11 s_d^2+2 s_d^4\right) s_g^2\right)\right)\biggr],\\
q_5&=&4 \sqrt{3} a^2 m^2 s_d c_df^2 \biggl[2 s_b c_b f^{-1} s_g \left(7 s_d^2+\left(-1+8 s_d^2+s_d^4\right) s_g^2\right)+4 s_b^3 c_b f^{-1} s_g \left(1+7 s_d^2+\left(-1+8 s_d^2+s_d^4\right) s_g^2\right)\nonumber\\
&&+f^{-2} c_g \left(s_d^2 \left(-1+s_d^2\right)+\left(-1-3 s_d^2+2 s_d^4\right) s_g^2\right)+2 s_b^2 c_g \left(-4-13 s_d^2+2 s_d^4+6 s_d^6+2 \left(1-14 s_d^2-s_d^4+6 s_d^6\right) s_g^2\right)\nonumber\\
&&+4 s_b^4 c_g \left(-4+s_g^2+s_d^2 \left(-8+s_d^2+3 s_d^4+\left(-14-s_d^2+6 s_d^4\right) s_g^2\right)\right)\biggr],\\
q_6&=&-2 \sqrt{3} a^3 m c_d f\left[\left(1+2 s_b^2\right) s_d^2 f^{-1} s_g+2 s_b c_b c_g\right],\\
q_7&=&16 \sqrt{3} a m^4 s_b c_d f^4\biggl[s_b s_d^6 f^{-2} \left(7+4 s_d^2\right) s_g c_g^2+c_b s_d^8 f^{-3} c_g^3\nonumber\\
&&+s_b^3 s_d^2 s_g \left(-20+18 s_d^2+148 s_d^4+149 s_d^6+42 s_d^8+\left(-8+38 s_d^2+159 s_d^4+151 s_d^6+42 s_d^8\right) s_g^2\right)\nonumber\\
&&-4 s_b^7 c_d^6 s_g \left(8+5 s_d^2+6 s_d^4+2 \left(4+3 \left(s_d^2+s_d^4\right)\right) s_g^2\right)+4 s_b^6 c_b c_d^6 f^{-1} c_g \left(-2-4 s_g^2+6 s_d^4 c_g^2-s_d^2 \left(1+6 s_g^2\right)\right)\nonumber\\
&&+s_b^2 c_b s_d^4 f^{-1} c_g \left(-3+13 s_g^2+18 s_d^6 c_g^2+s_d^2 \left(11+29 s_g^2\right)+s_d^4 \left(33+38 s_g^2\right)\right)\nonumber\\
&&+2 s_b^5 c_d^2 s_g \left(-16 c_g^2+12 s_d^8 c_g^2+s_d^4 \left(11+7 s_g^2\right)-4 s_d^2 \left(9+10 s_g^2\right)+s_d^6 \left(40+39 s_g^2\right)\right)\nonumber\\
&&+2 s_b^4 c_b c_d^2 f^{-1} c_g \left(-4 s_g^2+s_d^2 \left(-5-3 s_g^2+3 s_d^2 \left(s_g^2+2 s_d^2 f^{-2} c_g^2\right)\right)\right)\biggr],\\
q_8&=&8 \sqrt{3} a^2 m^3 s_d c_d f^3\biggl[2 s_b c_b s_d^4 \left(4+s_d^2\right) f^{-2} s_g c_g^2+s_d^6 f^{-3} c_g^3\nonumber\\
&&+s_b^2 s_d^2 f^{-1} c_g \left(-3+s_d^2+27 s_d^4+18 s_d^6+\left(17+11 s_d^2+26 s_d^4+18 s_d^6\right) s_g^2\right)\nonumber\\
&&+2 s_b^3 c_b s_g \left(-4+15 s_d^2+77 s_d^4+64 s_d^6+12 s_d^8+\left(4+21 s_d^2+76 s_d^4+63 s_d^6+12 s_d^8\right) s_g^2\right)\nonumber\\
&&+4 s_b^6 c_d^2 f^{-1} c_g \left(-2+3 s_g^2+6 s_d^6 c_g^2-s_d^2 \left(4+s_g^2\right)+s_d^4 \left(5+6 s_g^2\right)\right)\nonumber\\
&&+4 s_b^5 c_b s_d^2 c_d^2 s_g \left(23+25 s_g^2+6 s_d^4 c_g^2+s_d^2 \left(26+27 s_g^2\right)\right)\nonumber\\
&&+4 s_b^4 f^{-1} c_g \left(-1+4 s_g^2+s_d^2 \left(-5+10 s_g^2+s_d^2 \left(1+8 s_g^2+3 s_d^2 \left(5+3 s_d^2\right) c_g^2\right)\right)\right)\biggr],\\
q_9&=&4 \sqrt{3} a^3 m^2 c_d f^2 \biggl[2 s_d^4 f^{-2} s_g c_g^2-s_b c_b s_d^2 f^{-1} c_g \left(1-6 s_g^2+2 s_d^2 c_g^2\right)-2 s_b^3 c_b f^{-1} c_g \left(1+s_d^2 \left(1-6 s_g^2\right)+2 s_d^4 c_g^2\right)\nonumber\\
&&+2 s_b^4 s_g \left(-2+12 s_d^6 c_g^2-s_d^2 \left(5+6 s_g^2\right)+s_d^4 \left(15+14 s_g^2\right)\right)\nonumber\\
&&+s_b^2 s_g \left(-4+24 s_d^6 c_g^2-6 s_d^2 \left(1+2 s_g^2\right)+s_d^4 \left(33+28 s_g^2\right)\right)\biggr],\\
q_{10}&=&r_2a^4=-2 \sqrt{3} a^4 m s_d c_df \left[-2 s_b c_b s_g+f^{-1} c_g+2 s_b^2 f^{-1} c_g\right].\\
\end{eqnarray}

\section{D=4 SL(2,R) duality}
In this section, we summarize the results of the solution-generation-technique~\cite{MT} using the $SL(2,R)$ duality symmetry \cite{MO} 
of five-dimensional minimal supergravity dimensionally reduced to four dimensions. 
Since this solution-generation method is already described \cite{MT} in detail, 
we will be brief. 

The Lagrangian is
\beqa
{\cal L}
=E^{(5)}\left(R^{(5)}-\frac14 F_{MN}F^{MN}\right)
-\frac1{12\sqrt{3}}\epsilon^{MNPQR}F_{MN}F_{PQ}A_R.
\label{5DL}
\eeqa
$M,N,\ldots$ are five-dimensional curved indices running over $0,1,2,3$ and $5$.
$E^{(5)}$ is the determinant of the vielbein 
\beqa
E^{(5)A}_{~M}&=&\left(
\begin{array}{cc}
\rho^{-\frac12} E^{(4)\alpha}_{~~\mu}
&B_\mu \rho \\
0& \rho
\end{array}
\right)
\eeqa
of the metric 
$G^{(5)}_{MN}=E^{(5)A}_{~M}E^{(5)B}_{~N}\eta_{AB}$, 
$\eta_{AB}\equiv\mbox{diag}(-1,+1,+1,+1,+1)$.
$x^\mu$ ($\mu=0,1,2,3$) is the four-dimensional coordinates.
We take $\partial/\partial x^5$ as the Killing vector. 

\medskip

After the dimensional reduction and dualizing the gauge field $A_\mu$, we 
end up with a four-dimensional $SL(2,R)/U(1)$ non-linear sigma model coupled to 
two $U(1)$ gauge fields and gravity:
\beqa
{\cal L}+{\cal L}_{\rm Lag.mult.}
&=&E^{(4)}R^{(4)} + {\cal L}_S +{\cal L}_V~~~\mbox{(up to a complete square)},\n
{\cal L}_S
&\equiv&-E^{(4)}\left(\frac32\partial_{{\mu}}\ln \rho
                        \partial^{{\mu}}\ln \rho
                 +\frac12\rho^{-2}\partial_{{\mu}}A_5
                        \partial^{{\mu}}A_5\right),\n
{\cal L}_V
&\equiv&-\frac14E^{(4)}{\cal G}^T_{{\mu}{\nu}}
N^{{\mu}{\nu}{\rho}{\sigma}}
{\cal G}_{{\rho}{\sigma}}.
\label{4d}
\eeqa
Here $N^{{\mu}{\nu}{\rho}{\sigma}}$ is given by
\beqa
N^{{\mu}{\nu}{\rho}{\sigma}}
&=& m~{1}^{{\mu}{\nu}{\rho}{\sigma}}
+a~(\ast)^{{\mu}{\nu}{\rho}{\sigma}},\n
V^{-1}mV^{-1}&=&K-\frac12(\Phi\Phi^*K+K\Phi^*\Phi)
+\frac14\Phi\Phi^{*2}K\Phi,\n
V^{-1}aV^{-1}&=&-\Phi^*K-\Phi+\frac12(\Phi\Phi^{*2}K+K\Phi^{*2}\Phi)
 +\frac13\Phi\Phi^*\Phi-\frac14\Phi\Phi^{*3}K\Phi,
\eeqa
where 
${1}^{{\mu}{\nu}{\rho}{\sigma}}
\equiv \frac12\left(G^{(4){\mu}{\rho}}G^{(4){\nu}{\sigma}}
-G^{(4){\nu}{\rho}}G^{(4){\mu}{\sigma}}\right)$,
$(\ast)^{{\mu}{\nu}{\rho}{\sigma}}
\equiv\frac12 E^{(4)-1}\epsilon^{{\mu}{\nu}{\rho}{\sigma}}$,
with
\beqa
V \equiv \left(\begin{array}{cc}\rho^{-\frac12}& 0 \\
 0 &\rho^{\frac32}\end{array}
\right),~~
 \Phi\equiv\left(\begin{array}{cc}
 0 &\sqrt{3}\phi\\
 \sqrt{3}\phi& 0 \end{array}
\right),~~
\Phi^*\equiv\left(\begin{array}{cc} 2\phi& 0 \\ 0 &0\end{array}
\right),
~~
K \equiv(1+\Phi^{*2})^{-1},~ 
\phi\equiv\frac1{\sqrt{3}}\rho^{-1}A_5.
\eeqa
The two-component  vector 
${\cal G}_{{\mu}{\nu}}
=
\left(\begin{array}{c}\tilde{A}_{{\mu}{\nu}}\\
B_{{\mu}{\nu}}\end{array} \right)$
contains of the field strengths of two
$U(1)$ gauge fields $\tilde{A}_{{\mu}}$ and $B_\mu$,
where $\tilde{A}_{{\mu}}$ is a dual of $A_\mu$. Their relation is 
\beqa
\tilde{A}_{\mu\nu}&=&\rho(\ast F^{(4)})_{\mu\nu}-\frac 2{\sqrt 3}A_5 F^{(4)}_{\mu\nu}
+\frac1{\sqrt 3}A_5^2 B_{\mu\nu}\label{eq:tildeA1}
\eeqa
with 
$F^{(4)}_{{\mu}{\nu}}\equiv F'_{{\mu}{\nu}} + B_{{\mu}{\nu}}A_5$,
$F'_{{\mu}{\nu}}
\equiv\partial_{{\mu}}A'_{{\nu}}-\partial_{{\nu}}A'_{{\mu}}$ and 
$A'_{{\mu}}\equiv A_{{\mu}}-B_{{\mu}} A_5$.
%

Let 
$
{\cal F}_{\mu\nu}\equiv\left(
\begin{array}{c}
{\cal G}_{\mu\nu}\\
{\cal H}_{\mu\nu}
\end{array}
\right)$
be a four-component field strength vector consisting of ${\cal G}_{\mu\nu}$ and 
${\cal H}_{\mu\nu}~\equiv~\left(
\begin{array}{c}
{\cal H}_{\mu\nu}^{\tilde{A}}\\
{\cal H}_{\mu\nu}^B
\end{array}
\right)
\equiv m(\ast {\cal G})_{\mu\nu} -a~{\cal G}_{\mu\nu}$.
%
Also let ${\cal V}_-$, ${\cal V}_+$ be scalar-field dependent four-by-four matrices 
\beqa
{\cal V}_+=\left(
\begin{array}{cc}
~V~~&\\
&V^{-1}
\end{array}
\right),
\quad{\cal V}_-=\exp\left(
\begin{array}{cc}
&-\Phi^*\\
-\Phi&
\end{array}
\right).
\eeqa
Then
it was shown \cite{MO} that all the equations of motion and the Bianchi identity 
are invariant under the $SL(2,R)$ transformation
\beqa
{\cal F}_{\mu\nu}&\mapsto&\Lambda^{-1}{\cal F}_{\mu\nu},\label{eq:Ftr}\\
({\cal V}_-{\cal V}_+)^T{\cal V}_-{\cal V}_+ &\mapsto&
\Lambda^T({\cal V}_-{\cal V}_+)^T {\cal V}_-{\cal V}_+ \Lambda, \label{VVtoVVU-1}
\eeqa
where
$\Lambda$ is an $SL(2,R)$ group element 
generated by 
\beqa
&&
E'=\left(
\begin{array}{cccc}
 0 & 0 & 2 & 0 \\
 0 & 0 & 0 & 0 \\
 0 & \sqrt{3} & 0 & 0 \\
 \sqrt{3} & 0 & 0 & 0
\end{array}
\right),~~ 
F'=\left(
\begin{array}{cccc}
 0 & 0 & 0 & \sqrt{3} \\
 0 & 0 & \sqrt{3} & 0 \\
 2 & 0 & 0 & 0 \\
 0 & 0 & 0 & 0
\end{array}
\right)~ \mbox{and}~ H'=\left(
\begin{array}{cccc}
 1 & 0 & 0 & 0 \\
 0 & -3 & 0 & 0 \\
 0 & 0 & -1 & 0 \\
 0 & 0 & 0 & 3
\end{array}
\right).
\eeqa

Using this $SL(2,R)$ invariance, one can obtain a new solution by acting this 
$SL(2,R)$ transformation on various fields of a known solution. 
In this paper, $\Lambda$ is taken to be
\beqa
\Lambda&=&e^{-\alpha E'}e^{-\beta F'},
\eeqa
then the transformation 
rules are given by the following formulas:
\begin{eqnarray}
\rho^{new}=\frac{\rho}{\left(1+\alpha\beta+\beta \frac{A_5}{\sqrt{3}}\right)^2+\beta^2\rho^2},\label{eq:newrho}
\end{eqnarray}
\begin{eqnarray}
A_5^{new}=\sqrt{3}\frac{\left(\alpha+\frac{A_5}{\sqrt{3}}\right)\left(1+\alpha\beta+\beta \frac{A_5}{\sqrt{3}}\right)+\beta\rho^2}{\left(1+\alpha\beta+\beta \frac{A_5}{\sqrt{3}}\right)^2+\beta^2\rho^2},\label{eq:newA5}
\end{eqnarray}
\begin{eqnarray}
B_\mu^{new}=\sqrt{3}\beta^2\gamma \tilde A_\mu+\left(\gamma^3+\sqrt{3} \beta\gamma^2 A_5\right)B_\mu-\sqrt{3} \beta\gamma^2 A_\mu+\beta^3 \tilde B_\mu, \label{eq:Bnew1}
\end{eqnarray}
\begin{eqnarray}
A_\mu^{new}&=&\left[\sqrt{3}\beta^2\gamma A_5^{new}-\beta(2+3\alpha\beta) \right]\tilde A_\mu+\left[-\sqrt{3}\alpha\gamma^2+\gamma^3 A_5^{new}-A_5((1+4\alpha\beta+3\alpha^2\beta^2)-\sqrt{3}\beta\gamma^2 A_5^{new}))\right]B_\mu\nonumber\\
&&+\left[(1+4\alpha\beta+3\alpha^2\beta^2)-\sqrt{3}\beta\gamma^2 A_5^{new})\right] A_\mu+\left[-\sqrt{3}\beta^2+\beta^3 A_5^{new}\right]\tilde B_\mu \label{eq:Anew1},
\end{eqnarray}
where 
$\tilde A_\mu$ and $\tilde B_\mu$ are ``vector potentials" of $\tilde A_{\mu\nu}$ 
and ${\cal H}^B_{\mu\nu}$ satisfying
\beqa
(d\tilde A)_{\mu\nu}=\tilde A_{\mu\nu},~~~
(d\tilde B)_{\mu\nu}={\cal H}^B_{\mu\nu},\label{eq:tB}
\eeqa
respectively. $E^{(4)}$ remains unchanged through this transformation.

\end{document}